\documentclass[runningheads]{llncs}

\usepackage{subcaption}
\usepackage{eccvabbrv}
\usepackage{pifont}
\usepackage{multirow}
\usepackage{graphicx}
\usepackage{booktabs}
\usepackage{booktabs}
\usepackage[accsupp]{axessibility}  

\usepackage{hyperref}

\usepackage{orcidlink}
\usepackage{authblk}
\usepackage{caption}
\makeatletter
\renewcommand{\thanks}[1]{\footnotetext{#1}}
\makeatother

\begin{document}

\title{X-SG\texorpdfstring{$^2$}{2}S: Safe and Generalizable Gaussian Splatting with X-dimensional Watermarks}
\titlerunning{X-SG\texorpdfstring{$^2$}{2}S}

\author{Zihang Cheng\texorpdfstring{$^{\dagger}$}{*}\inst{1}\orcidlink{0009-0001-8415-8686} \and
Wentao Bao\texorpdfstring{$^{\dagger}$}{*}\thanks{\texorpdfstring{$\dagger$}{*} Equal contribution}\inst{2} \and
Huiping Zhuang\texorpdfstring{$^{\star}$}{*}\inst{1}\orcidlink{0000-0002-4612-5445} \and
Chun Li\inst{3}\orcidlink{0000-0002-2021-751X} \and
Xin Meng\inst{4}\orcidlink{0000-0003-2228-0587} \and
Ziqian Zeng\inst{1} \and
Cen Chen\inst{1}\orcidlink{0000-0003-1389-0148} \and
Ming Li\texorpdfstring{$^{\star}$}{**}\thanks{\texorpdfstring{$\star$}{**} Corresponding authors}\inst{5}\orcidlink{0000-0002-7852-0159} \and
F. Richard Yu\inst{6}
}
\authorrunning{Z. Cheng et al.}

\institute{South China University of Technology, Guangzhou Guangdong 511436, China \and  
Independent Researcher, USA \and
Shenzhen MSU–BIT University, Shenzhen Guangdong 518172, China \and
Peking University, Haidian District Beijing 100871, China \and
School of Artificial Intelligence, The Chinese University of Hong Kong, Shenzhen Guangdong 518172, China \and
Carleton University, 1125 Colonel By Drive, Ottawa, ON, K1S 5B6, Canada\\
\email{ftcldj@mail.scut.edu.cn}\qquad\email{ming.li@u.nus.edu}\qquad\email{hpzhuang@scut.edu.cn}
}

\maketitle
\enlargethispage{2\baselineskip}

\begin{abstract}
  3D Gaussian Splatting (3DGS) has been widely used in 3D reconstruction and 3D generation. However, the rapid adoption of 3D Gaussian Splatting raises growing concerns about information leakage and unauthorized use, urging the exploration of effective watermarking techniques. However, existing methods are limited by low capacity, fragility under geometric perturbations, and the infeasible requirement for costly fine-tuning or pipeline modifications, motivating the need for a generalizable, feed-forward framework capable of robust multi-modal embedding with minimal intrusion. In this paper, we propose a new framework X-SG$^2$S which can simultaneously inject 1D to 3D watermarks for copyright protection, while keeping the high fidelity of original 3DGS scenes. Specifically, we first split the watermarks into message patches. A self-adaptive gate is developed to select the injection positions of the watermark messages. Then, we use an XD (multi-dimensional) injection head to inject multi-modal messages into sorted 3DGS points. To restore watermarking messages, a learnable gate is developed to recognize the watermarked locations, from which our XD-extraction heads are used to restore hidden messages. X-SG$^2$S is the first framework to unify 1D-to-3D watermarking and enable simultaneous multi-modal watermark embedding in 3DGS, achieving this with minimal rendering interference and zero modifications to parameters or pipelines. Extensive experiments demonstrate that X-SG$^2$S effectively preserves consistency between the watermark and the original 3DGS, exhibits robustness against model degradation, and maintains accurate judgment capabilities. 
  \keywords{Copyright Preservation \and Multi Modal Watermarks \and 3DGS}
\end{abstract}

\section{Introduction}
\begin{figure}[h]
\centering
\includegraphics[width=\columnwidth]{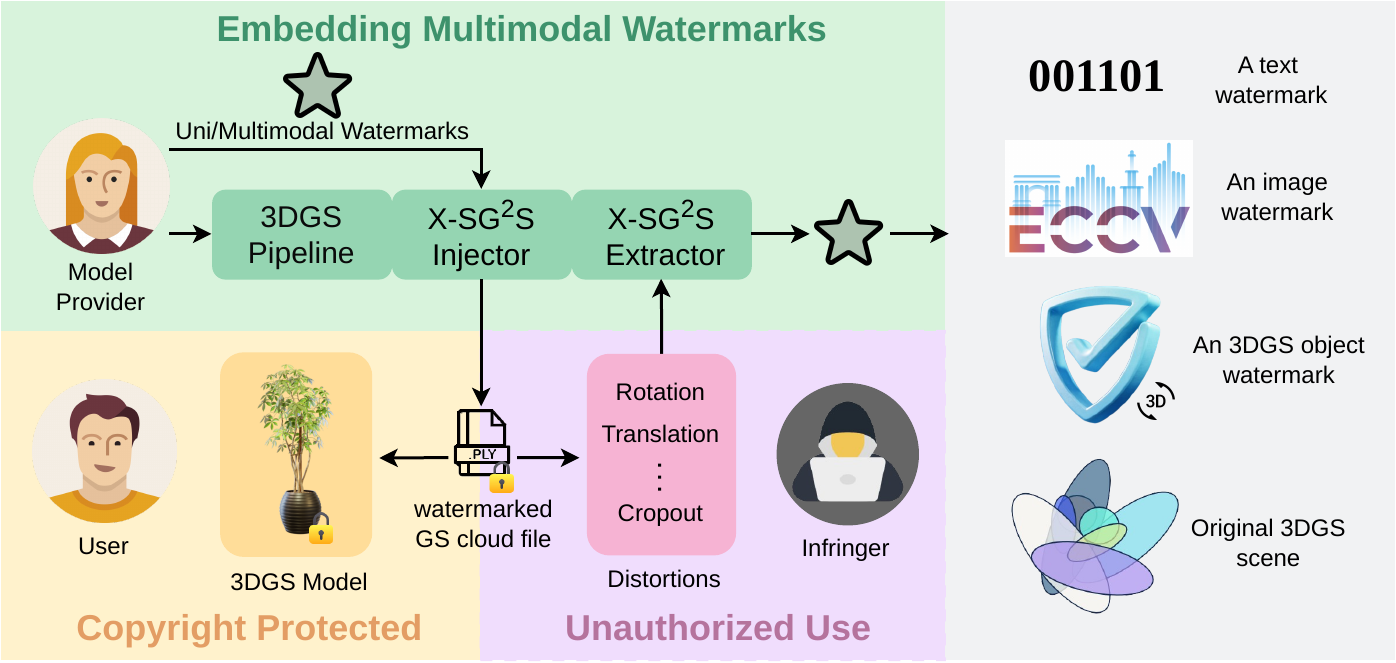}
\caption{Application Scenarios: 
After training, a 3DGS model can be paired with an X-SG$^2$S injector to embed multimodal watermarks e.g., text (“001101”), images (ECCV logo), or small 3D objects for branding. Outputs can later be verified via the extractor to confirm model usage.
2) The watermarks can be detectable even under various of hostile attacks.}
\end{figure}

With the growing demand for photorealistic 3D content, developing efficient and generalizable 3D representations has become a central research goal. Across domains such as healthcare, engineering, cultural heritage, VR/AR, and digital media, these techniques bridge the physical and digital worlds. Recent advances in 3DGS demonstrate its strong potential as an efficient and expressive representation for high-quality, real-time 3D reconstruction and rendering.

However, the rapid adoption of 3D Gaussian Splatting raises growing concerns about information leakage and unauthorized use which urgent to explore effective watermarking techniques. 
In practice, creators upload 3DGS assets to website platforms, where infringers can download, modify, and re-upload them, thereby violating the original creator’s rights. Thus, platforms require an invisible yet verifiable watermark that embeds rich multimodal metadata—such as creator ID or copyright protocol bits, QR codes, or 3D object brands logo. With these watermarks, platforms can be immediately notified when copyright is infringed. However, prior works only embed \textit{error-prone} and \textit{extremely low-capacity} bit watermarks into 3DGS and suffer degradation under geometric perturbations, posing security risks. Moreover, all these methods require fine-tuning the original model for watermark embedding, which is infeasible without multi-view images of the scene and demands significant computational cost and time~\cite{huang2024gaussianmarker,chen2025guardsplat,jang20253d}. Meanwhile, some watermarking methods alter the generation pipeline or the parameter structure of 3DGS, thereby limiting their generality~\cite{zhang2024gs,li2024gaussianstego}. A generalizable feed-forward multimodal watermarking framework can address these issues by enabling richer semantic embedding, and more robust, general verification with minimal intrusion to the 3DGS pipeline.

To bridge this gap, we propose X-SG$^2$S, a simple, effective, and flexible watermarking framework for 3D Gaussian Splatting. It is a feed-forward network, directly adding multimodal watermarks to partial spherical harmonic parameters and replacing the original which can avoid finetuning the model by using its training datasets. Meanwhile, our patch-wise and selective injection mechanism effectively preserves scene fidelity while isolating different modalities. Specifically, we first divide each type of watermarks into patches and introduce modality-specific redundancy  
to improve robustness during extraction. 
To enhance the robustness of image watermark recovery under partial corruption, we introduce a feature-sparse DCAE~\cite{chen2024deep} to reconstruct watermarked images from incomplete or noisy features. Next, we design a self-adaptive gate that learns to select optimal insertion points by modeling the interaction between the watermark content and the 3DGS representation. An X-dimensional (XD) injection head is used to embed each modality independently, while a learnable gate identifies and records the embedding positions. Finally, the embedded messages are retrieved using the corresponding XD extraction heads. 

X-SG$^2$S enables simultaneous embedding of 3D models, images, or binary messages into an existing 3DGS scene and supports accurate message extraction through a lightweight and non-intrusive process. Besides, it requires no fine-tuning or redesign of 3DGS pipeline, making it compatible with both feedforward architectures and iterative generative models such as diffusion models. 
Extensive experiments demonstrate that X-SG$^2$S can effectively embed multimodal watermarks without degrading rendering quality, while offering high fidelity, robustness, reliability, and flexibility. In a nutshell, the contributions and advantages of our X-SG$^2$S are summarized as follows:
\vspace{-2pt}
\begin{itemize}
\setlength{\itemsep}{0pt}
\item We propose \textbf{X-SG$^2$S}, a unified and feed-forward watermarking framework for 3DGS that completes watermark embedding within seconds.
\item X-SG$^2$S exhibits large capacity and strong versatility, that supports the simultaneous embedding of \textbf{1D} binary messages, \textbf{2D} images, and \textbf{3D} objects into a single 3DGS scene.
\item Our patchifying strategy and X-SG$^2$S backbone demonstrate robust security and high fidelity, preserving watermark-decoding quality even under a wide range of attacks.
\end{itemize}

\section{Related Work}

\subsection{Generalizable GS}
3D Gaussian Splatting ~\cite{kerbl20233d} has emerged as an efficient representation for high-quality 3D reconstruction, leveraging millions of anisotropic Gaussians for continuous and differentiable rendering. Building on this foundation, subsequent works have extended 3DGS to both object- and scene-level reconstruction. Methods such as Splatter Image~\cite{szymanowicz2024splatter} and SF3D~\cite{boss2024sf3d} focus on compact or category-specific objects, while MV-Splat~\cite{chen2025mvsplat} and PixelSplat~\cite{charatan2024pixelsplat} improve scalability and multi-view consistency for complex scenes. Beyond reconstruction, diffusion-based approaches like DiffGS~\cite{Zhou2024DiffGSFG} further integrate generative modeling, enabling image-to-3D, text-to-3D and point cloud to 3DGS generation. Collectively, these studies demonstrate the expanding versatility of 3DGS across reconstruction and generation tasks.

\subsection{Watermarking in Deep Learning}
Digital watermarking embeds imperceptible signals to verify copyright and protect multimedia ownership. Early image watermarking methods such as HiDDeN~\cite{zhu2018hidden} and HiNet~\cite{jing2021hinet} use CNNs to encode binary messages or watermark images while preserving visual quality, while VideoSeal~\cite{fernandez2024video} leverages temporal coherence for robust video watermarking. More recently, diffusion-based approaches like Videomark~\cite{hu2025videomark} and Safe-SD~\cite{ma2024safe} inject ownership signals into the stochastic noise trajectories of generative diffusion processes. These methods highlight watermarking’s dual role in safeguarding intellectual property and ensuring content traceability through imperceptible yet resilient signal embedding.

\subsection{3D Steganography and watermarking}
In 3D generative era \cite{li2026vlaattcadaptivetesttimecompute,li2026sentinelvlametacognitivevlamodel,fang2025your,cvpr/gait3d_v1,parsinggait_gps,He2023UnifiedSGGHOI,He2022OpenVocabularySGG}, some researches have made 3D steganography achieve good results in explicit geometry like meshes and point clouds~\cite{ohbuchi2002frequency,zhu2024rethinking,ferreira2020robust} or implicit geometer like NeRF~\cite{li2023steganerf,luo2023copyrnerf}. 3DGS is a new technology to represent explicit geometry. However, few works have been researched in steganography for 3DGS. GS-Hider~\cite{zhang2024gs} design a coupled secured feature attribute to replace the original 3DGS’s spherical harmonics coefficients and then use a scene decoder and a message decoder to disentangle the original RGB scene and the hidden message. Splat in Splat~\cite{guo2024splats} first train two scene by using the same location and shape of the cover scene and different color and opacity. Then they inject the secret scene into the cover one by an unlearnable method and use an AE to map the transparency variation of each point. This method can work only when you have the entire multi view datasets. And for every scene, you should learn a new AE which is very fussy. GaussianStego~\cite{li2024gaussianstego} uses DINO to add image watermarks and use U-Net to extract them via multi rendered views.  GaussianMarker~\cite{huang2024gaussianmarker} can only embed 48 bit watermarks. The added information might be insufficient in specific contexts.

\section{Method}
\begin{figure}[ht]
\centering
\centerline{\includegraphics[width=\textwidth]{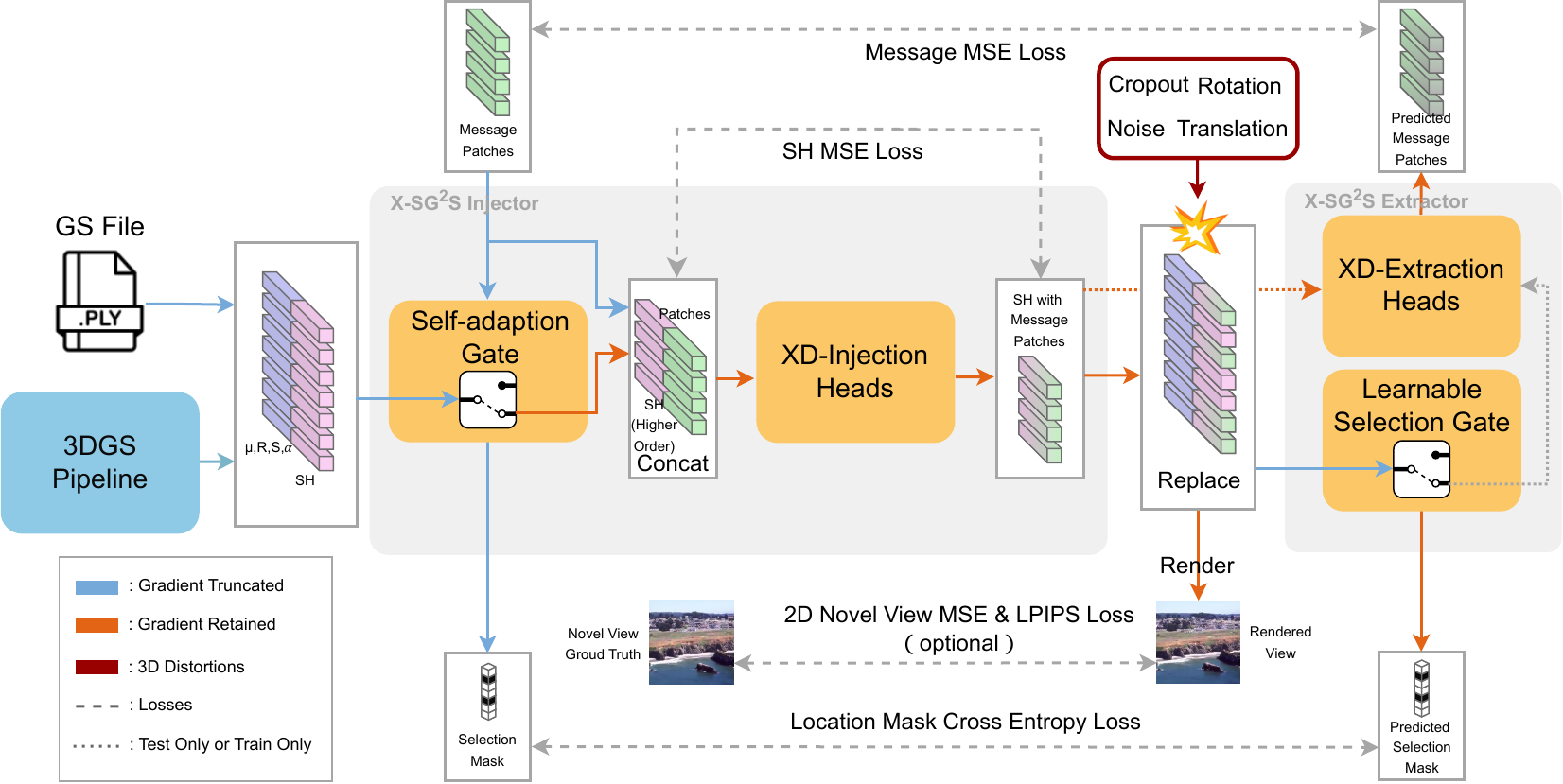}}
\caption{\textbf{Overview 
of X-SG$^2$S.} We first use patchify strategy splitting the watermark into small patches. At injection stage, adaptive selection gate  picks the 3DGS locations best suited for embedding given patches. XD-injection head embeds the patches into the high-order spherical-harmonic coefficients of the selected Gaussians. At extraction stage, a recognition gate retrieves the watermarked Gaussians and XD-extraction head decodes those patches. We truncate the gradient before the learnable selection gate and feed only the watermarked GS points under various 3D distortions to the XD-extraction head while training. 
}
\label{fig:train}
\end{figure}

\subsection{Proposed Method}
Our goal of 3D Gaussian Splatting watermarking is to inject multimodal watermarks into a 3DGS scene without compromising structure or rendering quality. 
The following objectives are desired in this task:
\begin{itemize}
    \item \textbf{3DGS Steganography:} embed auxiliary information into 3DGS scenes without damaging visual and structural fidelity.
    \item \textbf{Copyright protection:} embed user-defined watermarks for ownership verification, enabling detection of unauthorized usage by recovering injected watermarks. 
    \item \textbf{Multimodal embedding:} support text, image, and 3D object watermarks within one scene, enhancing the richness of the watermark while ensuring modality independence and accurate retrieval.
    \item \textbf{General and non-intrusive:} preserves the original 3DGS parameters and ensures seamless integration without any pipeline architectural modifications.
    \item \textbf{Fine-tuning free:} 
    no fine-tuning and no access to the images originally used to train the 3DGS model.
\end{itemize}
\subsection{Preliminary}
3D Gaussian Splatting ~\cite{kerbl20233d} is a rasterization-based representation enabling real-time, differentiable rendering of 3D scenes. A scene is represented as a set of anisotropic 3D Gaussians that collectively approximate both geometry and appearance.
\paragraph{Gaussian Representation.}
Each Gaussian $\mathcal{G}_i$ is defined by its center $\boldsymbol{\mu}_i \in \mathbb{R}^3$, covariance $\boldsymbol{\Sigma}_i \in \mathbb{R}^{3 \times 3}$ (parameterized by a scale vector $\boldsymbol{s}_i$ and a rotation quaternion $\boldsymbol{q}_i$), opacity $\alpha_i \in [0,1]$, and view-dependent color $\boldsymbol{c}_i(\boldsymbol{v})$ represented via Spherical Harmonics (SH) of order $k$.  
Its spatial density is:
\begin{equation}
    \rho_i(\boldsymbol{x}) = \exp\!\left[-\tfrac{1}{2}(\boldsymbol{x}-\boldsymbol{\mu}_i)^\top \boldsymbol{\Sigma}_i^{-1} (\boldsymbol{x}-\boldsymbol{\mu}_i)\right]
\end{equation}
A complete scene is thus $\mathcal{G} = \{(\boldsymbol{\mu}_i, \boldsymbol{\Sigma}_i, \alpha_i, \boldsymbol{c}_i)\}_{i=1}^N$.
\paragraph{Rasterization and Rendering.}
Given a camera with projection matrix $P$ and world-to-camera transform $W$, each Gaussian is projected to the image plane as
\begin{equation}
    \hat{\boldsymbol{\mu}}_i = PW\boldsymbol{\mu}_i~,\quad 
\hat{\boldsymbol{\Sigma}}_i = J_i \boldsymbol{\Sigma}_i J_i^\top
\end{equation}
where $J_i$ is the Jacobian of the local projection at $\boldsymbol{\mu}_i$.
The per-pixel contribution is
\begin{equation}
\sigma_i(\mathbf{p}) = \alpha_i \exp\!\left[-\tfrac{1}{2}(\mathbf{p}-\hat{\boldsymbol{\mu}}_i)^\top 
\hat{\boldsymbol{\Sigma}}_i^{-1}(\mathbf{p}-\hat{\boldsymbol{\mu}}_i)\right]    
\end{equation}
Final colors are accumulated via alpha compositing along the viewing ray:
\begin{equation}
    \mathbf{C}[\mathbf{p}] = 
\sum_{i=1}^{N} \boldsymbol{c}_i(\boldsymbol{v})\,\sigma_i(\mathbf{p})
\prod_{j<i}\!\left(1-\sigma_j(\mathbf{p})\right)
\end{equation}
This formulation enables efficient, differentiable, and photorealistic rendering of complex 3D scenes in real time.

\subsection{Methodology Overview}
\paragraph{Challenges.} 3DGS represents scenes as unordered point sets with rich parameters, posing unique challenges for watermarking: (1) Ordering invariance: the lack of inherent point order complicates the selection of points for watermarking, often necessitating global modifications~\cite{ohbuchi2002frequency} that waste capacity or degrade quality.
(2) Localization ambiguity: embedding in subsets of points lacks reliable indexing, especially under multiple watermarks.
(3) Limited per-point capacity: each Gaussian stores minimal information; efficiently encoding large payloads requires fragmentation and precise reconstruction strategies.
(4) Robustness issue: when certain points are missing or corrupted, watermark recovery must rely solely on the remaining or perturbed fragments which is usually challenging.
\paragraph{Solution.} To address these challenges, we develop X-SG$^2$S for 3DGS watermarking 
as depicted in Figure~\ref{fig:train}. We propose a modality-specific watermark splitting strategy. 
These patches are then processed by four modules that jointly inject and faithfully retrieve the watermark while keeping the original scene minimally perturbed: (1) an adaptive self-adaption gate for adaptively selecting GS points to use. (2) a multi-head MLP-based message injector for integrating information into SH parameters. (3) a learnable selection gate for recognizing the location of injected GS points. (4) a multi-head MLP-based message extractor for restoring message from the injected GS points. X-SG$^2$S injector is made up of (1) and (2), extractor is made up of (3) and (4). 

During training, the self-adaption gate generates a mask indicating the watermark injection locations and let the XD-injection head add the watermark patches. The learnable selection gate identifies those positions from the watermarked Gaussian parameters under various of 3D distortions and performs learning under the guidance of the mask. Simultaneously, the XD-extraction head extracts the patches from the Gaussian parameters watermarked by XD-injection head. During inference, only the Gaussian points selected by the learnable selection gate are sent to the XD-extraction head for watermark decoding.

\subsection{XD Watermark Representation}
To solve the challenges (3) and (4) introduced in the overview, watermarks are split into small patches with modality-aware redundancy. Patchfying reduces the information capacity per patch, avoiding overloading a single Gaussian point, which would not lead to information loss during recovery. 
Redundancy expansion further allows the watermarks to be reconstructed from their surviving counterparts when partial crops occur, yielding a robustness boost.

\begin{figure}[t]
\centering
\centerline{\includegraphics[width=\columnwidth]{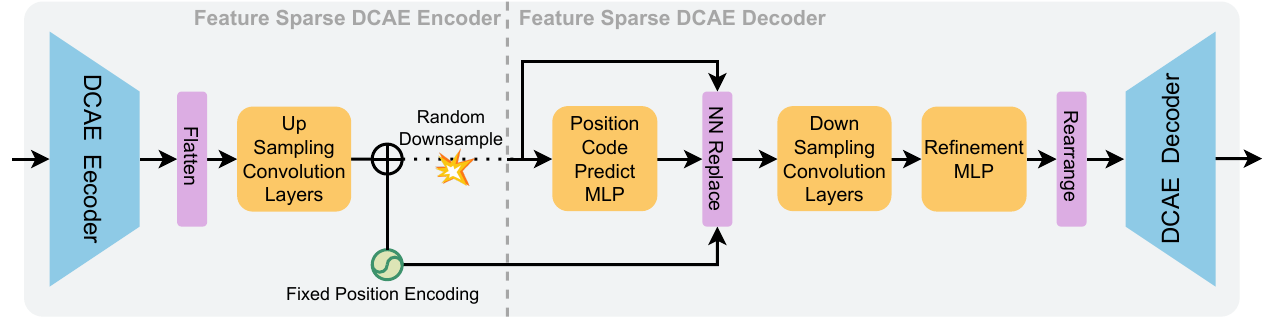}}
\caption{The structure of the feature-sparse DCAE.}
\label{fig:dcae}
\end{figure}

\paragraph{Patchifying Strategy.} We handle different modalities with modality-specific patching strategies. For 1D and 3D data, which are inherently discrete, we simply duplicate the original data multiple times to create redundancy. This allows accurate watermark recovery from any randomly sampled subset, ensuring robustness without additional processing. 
Features encoded by DCAE are split to patches and augmented with (i) position encoding and (ii) a compact summary of its neighboring patches, so that every token knows “where it is” in the feature map and the missing ones can be reconstructed from the survivors when partial corruptions occur.
In the Figure~\ref{fig:dcae}, we show our \textbf{feature-sparse DCAE}. DCAE~\cite{chen2024deep} encodes an image of resolution $n$ into a feature map of $\frac{n}{64}\times\frac{n}{64}\times d$, where the feature dimension is $d$. Then, the feature maps are flattened as a sequence with the shape of $\frac{n^2}{64^2}\times d$. Four up-sampling convolution layers are used to create redundant context for this sequence, with fixed position encoding added. 
To simulate a hostile attack, we randomly remove some patches. 
Benefited by the context sequence, our module is capable of restoring the whole sequence by any subset. Specifically, feature-sparse DCAE uses an MLP to predict the fixed position encoding. Then, it performs the nearest-neighbor search by comparing the predicted encoding and the complete encoding, which determines the patch locations and let all the patches be placed in the correct position. Then, down-sampling convolution layers and a refinement MLP are used to restore the original feature sequence. Finally, we rearrange the feature sequence and decode the watermark. More details about feature-sparse DCAE are introduced in the Appendix.
\paragraph{Message Embedding.} To enable uniform interaction, we use separate linear layers to map 1D, 2D, and 3D messages into a higher dimension latent space 
which provides rich information for computing watermarking locations.

\subsection{Self-Adaptive Selection of Injection Position}
Selecting optimal 3DGS points for message injection while preserving scene fidelity is non-trivial in practice. 
To tackle this challenge, we consider the point selection from two perspectives: 
the ``self-score'' and the ``cross-score''.
The ``self-score'' measures how much a point contributes to the overall 3DGS appearance. The ``cross-score'' reflects the alignment between the embedded message and the point cloud structure. 
Eventually, a top-$k$ mask from the product of the two scores can be used to determine where to inject watermarks. In light of the two scores, we develop a self-adaptation gate, as illustrated in Figure~\ref{fig:stru}, which will be entailed below.


\noindent\textbf{Self-score.} Computing the self-score requires each point to interact with the entire 3DGS. But vanilla Transformer incurs a prohibitive memory cost. To reduce memory complexity while retaining the interaction capability, we adopt Induced Set Attention~\cite{lee2019set}, which is efficient and order-invariant. Given $n$ Gaussian point embeddings $X \in \mathbb{R}^{n \times d}$ and $m$ learnable inducing tokens $I \in \mathbb{R}^{m \times d}$, we compute
\begin{equation}
\begin{aligned}
& I^*                 = f(I, X) \\
& O_s               = f(X, I^*) \\
& \text{self-score} = \sigma\bigl(h(O_s)\bigr)
\end{aligned}    
\end{equation}
where $f(\cdot, \cdot)$ is multi head attention block and $h(\cdot)$ is a linear layer. We use $m = 1$ and one attention head, reducing complexity from $O(n^2)$ to $O(n)$.

\noindent\textbf{Cross-score.} To compute the ``cross-score'', we use Efficient Attention~\cite{shen2021efficient}. Efficient attention (EA) not only drastically cuts computation but also preserves rich interaction between the 3DGS and watermarks. 
Given $l$ message patches $Y \in \mathbb{R}^{l \times d}$, the attention output is
\begin{equation}
    \begin{aligned}
& O_c                 = \sigma_{\text{row}}(X) \left[ \sigma_{\text{col}}(Y)^\top Y \right] \\
& \text{cross-score} = \sigma(h(O_c))
\end{aligned}
\end{equation}
Finally, the gate score is obtained by multiplying the ``self-score'' and ``cross-score'' element-wise.
Table~\ref{s} shows the efficiency of our self-adaptive gate.

 
\subsection{Learnable Selection Gate}
The previous self-adaptation gate can produce a mask when injecting a watermark into a 3DGS scene. However, another equally important task is to recover the watermark from the injected positions, which has to be learned by a module.  
We treat the predicted masks from the self-adaptation gate as the ground truth. The learnable selection gate regards the point set with added watermarks as input to find out the locations where the information is added. The gate is a four-layer MLP to predict the true location. 
During training, we detach the gradients before the learnable selection gate, ensuring it only learns where to embed the watermark.


\subsection{XD-Injection/Extraction Heads}
XD-Injection modules embed and extraction heads extract watermark patches. For each modality, we use a separate MLP following PointNet~\cite{qi2017pointnet}, typically with 3–5 layers. We select GS points from target locations suggested by the self-adaptive gate and extract their high-order spherical harmonic (SH) parameters from three channels. These are concatenated with the message patches and passed through a multi-head injector to produce SH parameters carrying the embedded information.
And for the extractor, it takes the SH parameters of injected points found out by the learnable selection gate as input and extracts the messages from them. The extracted messages will be optimal by the prediction loss as well. 
\begin{figure}[t]
\centering
\centerline{\includegraphics[width=\columnwidth]{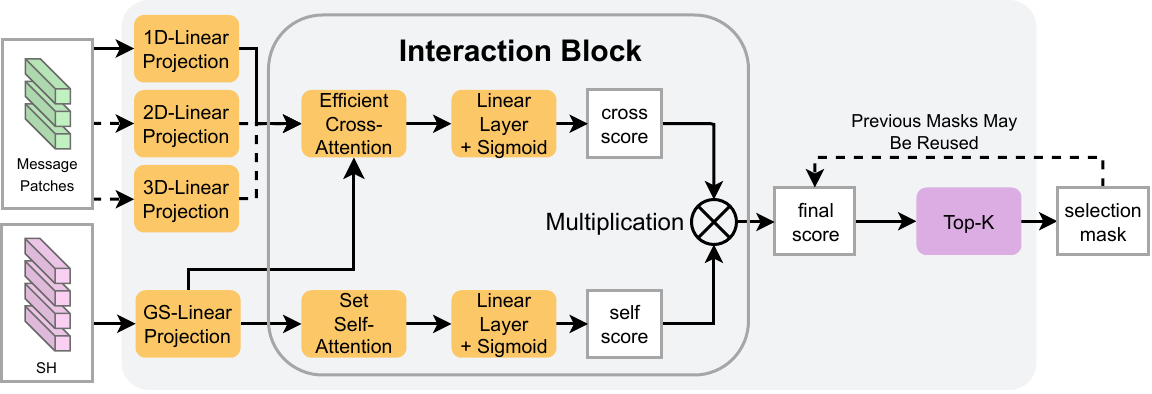}}
\caption{The structure of the self-adaptation gate.}
\label{fig:stru}
\end{figure}

\subsection{Training Loss}
To effectively train X-SG$^2$S, we develop a comprehensive loss function comprising four components: message patches loss, location mask loss, SH parameters loss, and other optional perceptual losses. 
\begin{equation}
    \begin{split}
    LOSS = \phi\mathcal{L}_\text{1D-BCE} + \theta\mathcal{L}_\text{2D-MSE} + \delta\mathcal{L}_\text{3D-MSE} \\
    + \mathcal{L}_\text{Mask-BCE} + \gamma\mathcal{L}_\text{SH-MSE} + \mathcal{L}_{\text{Optional}}
    \end{split}
\end{equation}
The message patches loss enforces accurate recovery of the injected information by ensuring consistency between the embedded and extracted messages. This is fulfilled by a cross-entropy loss $\mathcal{L}_\text{1D-BCE}$ for 1D messages and mean squared error (MSE) losses $\mathcal{L}_\text{2D-MSE}$ and $\mathcal{L}_\text{3D-MSE}$ for 2D and 3D messages, respectively. Mask cross-entropy loss $\mathcal{L}_\text{Mask-BCE}$ guides the learnable selection gate to precisely identify regions for message embedding. The SH parameters loss penalizes large deviations in spherical harmonic (SH) coefficients through an MSE loss $\mathcal{L}_\text{SH-MSE}$. Additionally, other losses, such as 2D view reconstruction loss and LPIPS loss, can be incorporated when leveraging pretrained 3DGS pipelines to further enhance the preservation of SH parameters and overall rendering fidelity.

\section{Experiments}
\subsection{Experimental Setup}
\textbf{Datasets.} We use a pretrained model MVSplat for real time inference. We use large-scale ACID~\cite{Liu2020InfiniteNP} datasets to let MVSplat~\cite{chen2025mvsplat} generate the 3DGS scenes. For 1D dataset, we randomly generate a sequence of binary message. For each bit, it has the same probability to be 1 or 0. For 2D dataset, we use Logo-2K~\cite{wang2020logo} dataset. For 3D dataset, we first download a subset of Objaverse~\cite{deitke2023objaverse}, and use Gamba~\cite{shen2024gamba} to generate the 3DGS objects. They are then saved in .ply format as a 3D dataset.

Specifically, ACID contains nature scenes captured by aerial drones, which are split into 11,075 training scenes and 1,972 testing scenes. After reconstructing Logo-2K dataset, it has 110313 training logos and 28415 testing logos. 3D dataset made by objaverse and Gamba contains 353 training files and 94 testing files. To ensure the number of datasets is aligned, we reuse the GS cloud files while training.

\noindent\textbf{Metrics.}
For original GS scenes and GS objects, we use pixel-level PSNR, patch-level SSIM~\cite{wang2004image}, and feature-level LPIPS~\cite{zhang2018unreasonable} to measure the completeness of the original scenes and the quality of restoration of the GS objects. For a fair comparison, all the scenes are rendered to 256$\times$256 resolutions while all the objects are rendered to 512$\times$512 resolutions. For 1D data, we use precision rate to measure the extraction accuracy of binary data. For 2D data, we use pixel-level PSNR to quantization the restoration of the 2D images.

\subsection{Implementation Details}
X-SG$^2$S is implemented with PyTorch, along with an off-the-shelf 3DGS render implemented in CUDA. All models are trained on a single RTX A6000ada with the Adam optimizer. 
The optional loss functions include 2D rendered image MSE loss and LPIPS loss, with weights of 1 and 0.05, respectively. And other hyperparameter are $\gamma = 1$, $\phi = 0.005$, $\theta = 2$, $\delta = 1.5$. We choose a binary message of 48bit$\times$1024 as a 1D watermark and a 3DGS object with 10,000 points as a 3D watermark. For the 2D watermark, we use feature sparse DCAE to get 16$\times$2048 message patches. The number of higher-order SH parameters is set to 48 for each GS point.

\begin{figure}[t]
  \centering
  \begin{subfigure}[b]{.16\linewidth}
    \includegraphics[width=\linewidth]{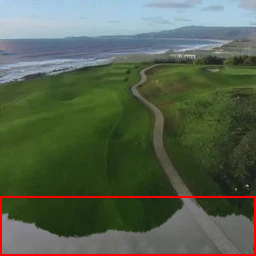}
  \end{subfigure}
  \begin{subfigure}[b]{.16\linewidth}
    \includegraphics[width=\linewidth]{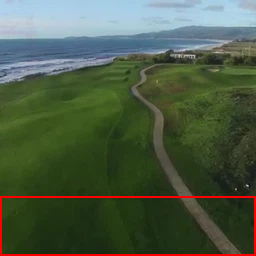}
  \end{subfigure}
  \begin{subfigure}[b]{.16\linewidth}
    \includegraphics[width=\linewidth]{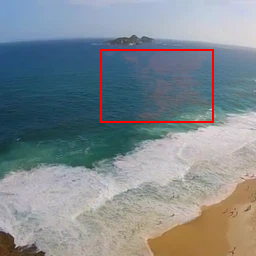}
  \end{subfigure}
    \begin{subfigure}[b]{.16\linewidth}
    \includegraphics[width=\linewidth]{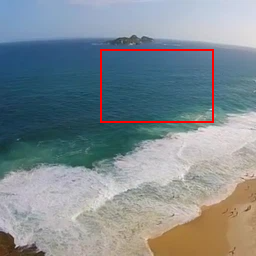}
  \end{subfigure}
  \begin{subfigure}[b]{.16\linewidth}
    \includegraphics[width=\linewidth]{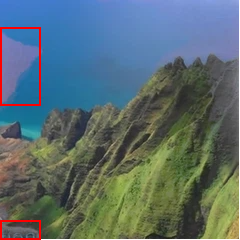}
  \end{subfigure}
  \begin{subfigure}[b]{.16\linewidth}
    \includegraphics[width=\linewidth]{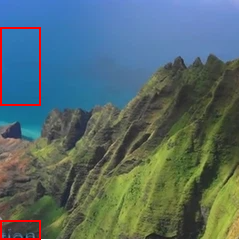}
  \end{subfigure}
  \caption{This image illustrates the different effects of watermarking all SH coefficients versus watermarking only the high-order SH coefficients. The odd columns show X-SG$^2$S injects the watermark to all the SH while the even columns show the watermarks are only injected to the higher order SH.}
  \label{fig:higher}
\end{figure}

\begin{figure}[!ht]
  \centering
  \begin{subfigure}[b]{.16\linewidth}
    \includegraphics[width=\linewidth]{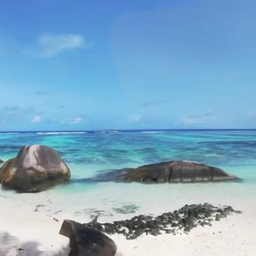}
  \end{subfigure}
  \begin{subfigure}[b]{.16\linewidth}
    \includegraphics[width=\linewidth]{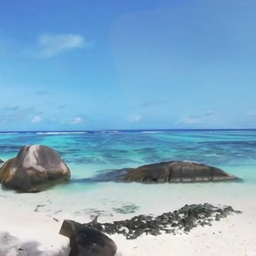}
  \end{subfigure}
  \begin{subfigure}[b]{.16\linewidth}
    \includegraphics[width=\linewidth]{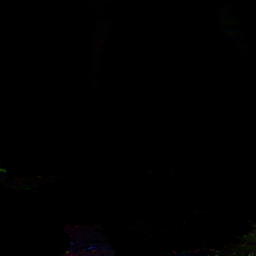}
  \end{subfigure}
  \begin{subfigure}[b]{.16\linewidth}
    \includegraphics[width=\linewidth]{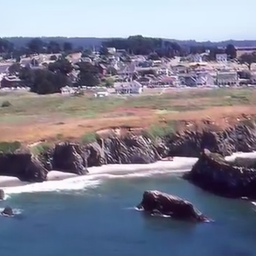}
  \end{subfigure}
  \begin{subfigure}[b]{.16\linewidth}
    \includegraphics[width=\linewidth]{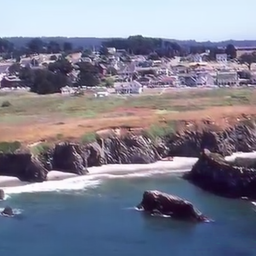}
  \end{subfigure}
  \begin{subfigure}[b]{.16\linewidth}
    \includegraphics[width=\linewidth]{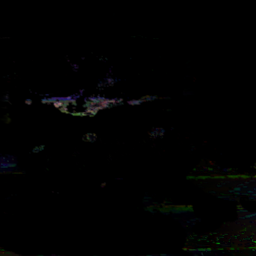}
  \end{subfigure}

  \caption{This image demonstrates the effectiveness of our approach. The first column shows the original scene images rendered by MVSplat. The second column displays the images after watermark embedding. The third column presents the 10×residuals between the two images. It can be seen that our model is capable of effectively concealing the watermark.}
  \label{fig:res}
\end{figure}

\begin{table}[t]
\centering
\scriptsize
\setlength{\tabcolsep}{1.5pt}
\caption{Comparison of the quantitative performance of the original scenes and watermarks under various perturbations on ACID~\cite{Liu2020InfiniteNP} and Mip-NeRF360~\cite{barron2022mip} dataset.}
\label{r}
\begin{tabular}{l ccccc}
\toprule
\textbf{Attack Type} & 
\multicolumn{5}{c}{\textbf{Model Ability}} \\
 \cmidrule(lr){2-6}
  & Acc & PSNR (image) & PSNR (3D obj) & SSIM (3D obj) & LPIPS$\downarrow$(3D obj) \\
\midrule
None                         & 100   & 22.791 & 23.017 & 0.836 & 0.138 \\
Noise to SH($\sigma=0.05$)   & 96.22 & 20.008 & 20.307 & 0.796 & 0.219 \\
Translation($t=[0, 1000]^3$) & 99.31 & 22.669 & 23.010 & 0.835 & 0.139 \\
Rotation($r=\pm\pi/6$)       & 98.56 & 22.139 & 22.718 & 0.834 & 0.141 \\
Cropout($cr=0.1$)            & 100   & 22.539 & 23.005 & 0.835 & 0.139 \\

\bottomrule
\end{tabular}
\end{table}

\begin{table}[t]
\centering
\scriptsize
\setlength{\tabcolsep}{0.8pt}
\caption{Comparison of bit accuracy under various perturbations on Blender~\cite{mildenhall2021nerf} and LLFF~\cite{mildenhall2019local} datasets. We also report the watermarking speed and watermarking capability. We show the results on 48-bit messages. Bold indicate the best performance.}
\begin{tabular}{@{}l c c ccccc@{}}
\toprule
\textbf{Method} & 
\multicolumn{2}{c}{\textbf{Model Ability}} &
\multicolumn{5}{c}{\textbf{Bit Acc} $\uparrow$ (\%)} \\
 \cmidrule(lr){2-3} \cmidrule(lr){4-8}
 & \shortstack{MM \\ wms} & wm spd
 & \shortstack{None} & \shortstack{Noise to SH\\($\sigma=0.05$)} & \shortstack{Translation\\($t=[0,10^3]^3$)} & \shortstack{Rotation\\($r=\pm\pi/6$)} & \shortstack{Cropout\\($cr=0.1$)}\\
\midrule
3DGS~\cite{kerbl20233d} w/ fine-tuning                   & \ding{55} & 3-5 mins          & 69.79          & 69.70  & 68.78 & 65.88 & 64.84 \\
GaussianMarker~\cite{huang2024gaussianmarker} 2D dec & \ding{55} & 3-5 mins          & 97.85          & 57.05  & 59.07 & 53.88 & 48.23 \\
GaussianMarker~\cite{huang2024gaussianmarker} 3D dec & \ding{55} & 3-5 mins          & \textbf{100}   & \textbf{99.90}  & 98.95 & 95.83 & 92.70 \\
GuardSplat~\cite{chen2025guardsplat}                     & \ding{55} & 3-5 mins          & 90.98          & 67.10  & 67.77 & 66.13 & 71.32 \\
3D-GSW~\cite{jang20253d}                                 & \ding{55} & 3-5 mins          & 93.72          & 84.54  & 81.67 & 82.33 & 91.87 \\
Ours                                                     & \ding{51} & \textbf{3 secs} & \textbf{100}   & 94.12 & \textbf{99.07} & \textbf{96.56} & \textbf{95.01} \\
\bottomrule
\end{tabular}
\label{comparebl}
\end{table}

\subsection{Fidelity}
\begin{figure}[htb]
\centering
\begin{minipage}{0.65\linewidth}
    \centering
      \begin{subfigure}[b]{.22\linewidth}
        \includegraphics[width=\linewidth]{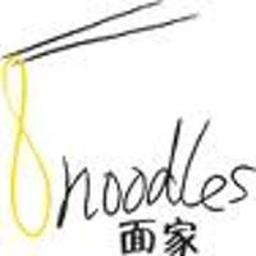}
      \end{subfigure}
      \begin{subfigure}[b]{.22\linewidth}
        \includegraphics[width=\linewidth]{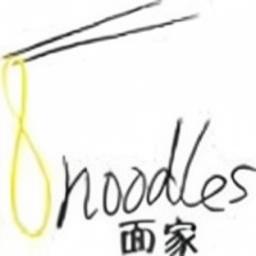}
      \end{subfigure}
      \begin{subfigure}[b]{.22\linewidth}
        \includegraphics[width=\linewidth]{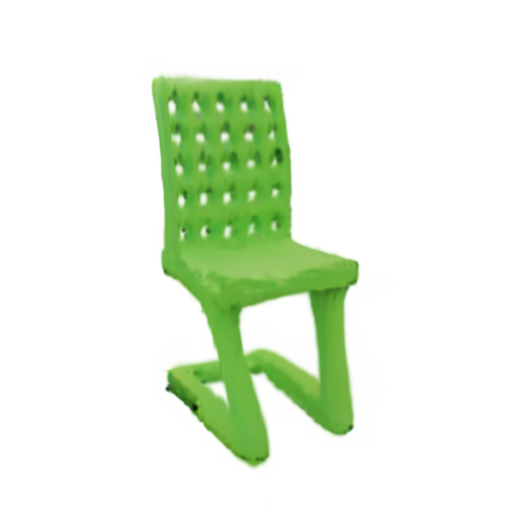}
      \end{subfigure}
      \begin{subfigure}[b]{.22\linewidth}
        \includegraphics[width=\linewidth]{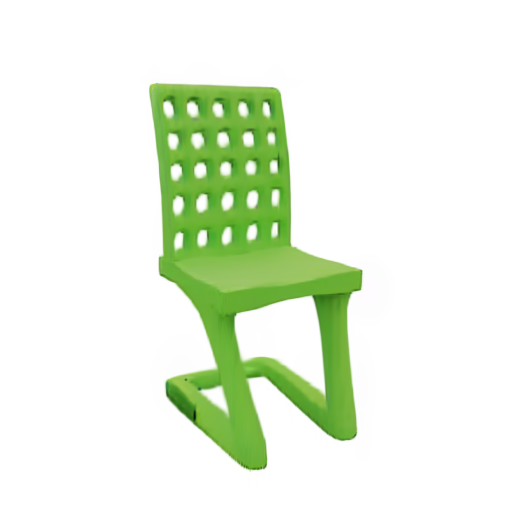}
      \end{subfigure}
      
      
      \begin{subfigure}[b]{.22\linewidth}
        \includegraphics[width=\linewidth]{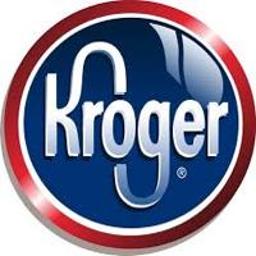}
      \end{subfigure}
      \begin{subfigure}[b]{.22\linewidth}
        \includegraphics[width=\linewidth]{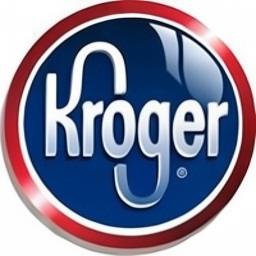}
      \end{subfigure}
      \begin{subfigure}[b]{.22\linewidth}
        \includegraphics[width=\linewidth]{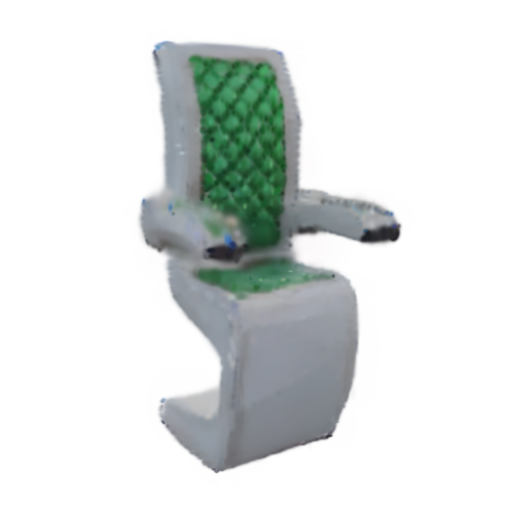}
      \end{subfigure}
      \begin{subfigure}[b]{.22\linewidth}
        \includegraphics[width=\linewidth]{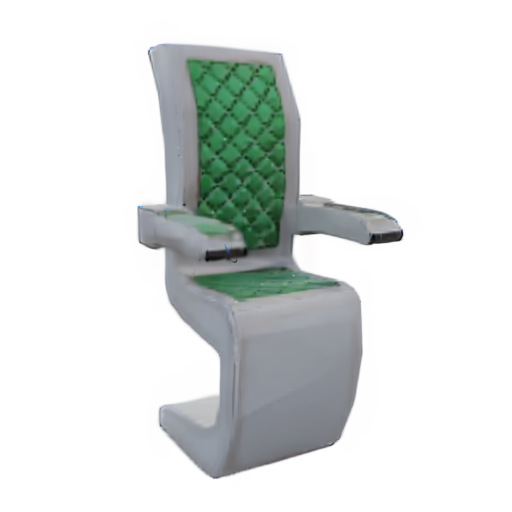}
      \end{subfigure}
\end{minipage}
\caption{The image shows watermarks extracted from 3DGS scenes by X-SG$^2$S. The first column shows the original image watermarks provide by Logo-2K~\cite{wang2020logo}. The second column displays the image watermarks extracted by X-SG$^2$S. The third column displays the GS objects provided by Gamba~\cite{shen2024gamba}. The final column presents the GS objects extracted by X-SG$^2$S.}
\label{fig:wm}
\end{figure}

First, we evaluate where to embed the watermark patches within the spherical harmonic (SH) coefficients of 3DGS can minimize the distortion to the original scene. To this end, we conduct an experiment by injecting identical watermarks into all SH coefficients and  only into the higher-order SH coefficients separately. According to figure~\ref{fig:higher}, we show the importance of SH. It seems that only inject watermarks to higher order SH can achieve better results. At the same time, when only adding watermarks to higher order SH, the watermarks cannot be detected by the naked eye from the visual perspective which means we protect the original scene well and hide the watermarks well. 
Meanwhile,we have drawn the residual map of the invisible watermark in figure~\ref{fig:res}. It can be observed that our method can effectively hide the watermark without detection. 
In figure~\ref{fig:wm}, we also shows 2D and 3D watermark extracted by our X-SG$^2$S. The decoded watermarks are clearly legible and of high quality which means our X-SG$^2$S can inject and extract watermarks well.

\subsection{Robustness}
We have subjected the Gaussians to 3D degradation using random pruning, rotation, translation adding gaussian noise methods. The following distortions for message extraction are considered: Gaussian Noise($\sigma$ = 0.1), Rotation(random select between +$\pi/6$ and -$\pi/6$), Translation(random select between $[0, 1000]^3$), Cropout($10\%$ of the original). Quantitative metrics are shown in Table~\ref{r}. The results indicate that our method effectively withstands the degradation process.

\subsection{Compare with Other Methods}
We compare X-SG$^2$S with 5 baselines on binary watermark to guarantee a fair comparison: (1) 3DGS~\cite{kerbl20233d} with fine-tuning, (2) GaussianMarker 2D Decoder~\cite{huang2024gaussianmarker}, (3) GaussianMarker 3D Decoder~\cite{huang2024gaussianmarker}, (4) GuardSplat~\cite{chen2025guardsplat} and (5) 3D-GSW~\cite{jang20253d}. We embed 48-bit watermarks into the same  scenes using these different methods. Table~\ref{comparebl} demonstrate that our approach effectively withstands external 3D perturbations and achieves state-of-the-art performance.

\subsection{Precise Identification}
In the real case, we cannot indiscriminately detect watermarks. That is to say, we cannot identify a watermark in a model that does not have one. To illustrate this, we randomly select some scenes without watermarks and observe that our model fails to detect any watermark. Table~\ref{d} demonstrates that our XD extraction head can not extract watermark amount unwatermarked 3DGS model which can effectively reduce false positives without the need for an additional discriminator. 

\subsection{Ablation Results}
\paragraph{The Effectiveness Of Our Interaction Block.}
We design three different structures of interaction block and conduct ablation experiments respectively: (1) ``Interactive": enabling interactions between Gaussian points and watermark blocks as shown in Figure~\ref{fig:stru}. (2) ``Normal": using 3DGS original points as input and passing them through n layers of MLP gate. (3) ``Random": randomly selecting k locations to add an extra message that can not be learned.  Table~\ref{s} demonstrates that enabling interactions between points and watermarks significantly improves performance.

\begin{table}[t]
    \centering
    \begin{minipage}{0.49\linewidth}
        \center
        \tiny
        \setlength{\tabcolsep}{5pt}
        \caption{To verify whether the detection model can distinguish if the original model contains a watermark. Results show X-SG$^2$S can unambiguously determine.}
        \begin{tabular}{lcc}
        \toprule
                                  & GS w/o wm & GS w wm      \\
        \midrule
        PSNR (org)                 & 27.826    & 27.804      \\
        SSIM (org)                 & 0.871     & 0.871       \\
        LPIPS (org)$\downarrow$    & 0.123     & 0.126       \\
        Acc                       & 0.395     & 1.000       \\
        PSNR (image)               & 7.719     & 22.791      \\
        PSNR (3D obj)              & 11.934    & 23.017      \\
        SSIM (3D obj)              & 0.835     & 0.836       \\
        LPIPS (3D obj)$\downarrow$ & 0.319     & 0.138       \\
        \bottomrule
        \end{tabular}
        \label{d}
    \end{minipage}
    \hfill 
    \begin{minipage}{0.49\linewidth}
        \center
        \tiny
        \setlength{\tabcolsep}{2.5pt}
        \caption{Single/Multi-modality watermark injection and extraction. X-SG$^2$S supports watermark injection for both single-modal and multi-modal data.}
        \begin{tabular}{lccccr}
        \toprule
        Method & 1D & 2D & 3D & XD & Org  \\
        \midrule
        PSNR(org)                & 27.814         & 27.811         & 27.810      & 27.804      & 27.826  \\
        SSIM(org)                & 0.871          & 0.871          & 0.871       & 0.871       & 0.871  \\
        LPIPS(org)$\downarrow$   & 0.124          & 0.125          & 0.125       & 0.126       & 0.123  \\
        Acc                      & 1.000          & /              &  /          & 1.000       & 1.000    \\
        PSNR(image)              & /              & 22.804         &  /          & 22.791      & 29.755    \\
        PSNR(3D obj)             & /              & /              & 23.044      & 23.017      & 28.807 \\
        SSIM(3D obj)             & /              & /              & 0.840       & 0.836       & 0.942  \\
        LPIPS(3D obj)$\downarrow$& /              & /              & 0.137       & 0.138       & 0.067  \\
        \bottomrule
        \end{tabular}
        \label{single}
    \end{minipage}
\end{table}

\paragraph{Single Modality Injection Test.}
In Table~\ref{single}, we test our X-SG$^2$S on adding sigle watermark. Because the different modalities are mutually non-interfering during embedding, our model can flexibly inject either single-modal or multi-modal watermarks while guaranteeing their exact recovery. We can see that X-SG$^2$S can well adapt the task of 1 to 3D watermarking and the performance of watermark extraction are almost identical.
\paragraph{The Exploration of Loss Function.}
Here we set the weights of optional losses to 0. From Table~\ref{noloss}, we can see that this way is feasible. However, X-SG$^2$S trained without optional loss performs worse. Therefore, we recommend using optional loss functions to enhance the effectiveness of the training.
\begin{table}[t]
    \centering
    \begin{minipage}{0.49\linewidth}
        \center
        \tiny
        \setlength{\tabcolsep}{1pt}
        \caption{Comparison with and without optional loss functions. Results show X-SG$^2$S trained with optional loss performs better.}
        \begin{tabular}{lcccr}
        \toprule
        Method & w/o optional losses & w optional losses  \\
        \midrule
        PSNR (org)                & 27.757        & \textbf{27.804}     \\
        SSIM (org)                & 0.869         & \textbf{0.871}      \\
        LPIPS (org)$\downarrow$   & 0.128         & \textbf{0.126}      \\
        Acc                      & 0.979         & \textbf{1.000}        \\
        PSNR (image)              & 21.923        & \textbf{22.791}    \\
        PSNR (3D obj)             & 22.647        & \textbf{23.017}    \\
        SSIM (3D obj)             & 0.833         & \textbf{0.836}     \\
        LPIPS (3D obj)$\downarrow$& 0.140         & \textbf{0.138}     \\
        \bottomrule
        \end{tabular}
        \label{noloss}
    \end{minipage}
    \hfill 
    \begin{minipage}{0.49\linewidth}
        \center
        \tiny
        \setlength{\tabcolsep}{2.5pt}
        \caption{Comparing different structures of interaction block. Our module adaptively aligns 3DGS and watermarks to optimize embedding locations.}
        \begin{tabular}{lccccr}
        \toprule
        Method                   & Ours     & Normal      & Random   & Org      \\
        \midrule
        PSNR (org)                & \textbf{27.804}     & 24.276      & 21.534 & 27.826    \\
        SSIM (org)                & \textbf{0.871}      & 0.824       & 0.529  & 0.871    \\
        LPIPS (org)$\downarrow$   & \textbf{0.126}      & 0.235       & 0.443  & 0.123    \\
        Acc                      & \textbf{1.000}      & 0.958       & 0.667  & 1.000    \\
        PSNR (image)              & \textbf{22.791}     & 22.595      & 21.797 & 29.755   \\
        PSNR (3D obj)             & \textbf{23.017}     & 21.322      & 20.699 & 28.807   \\
        SSIM (3D obj)             & \textbf{0.836}      & 0.833       & 0.831  & 0.942    \\
        LPIPS (3D obj)$\downarrow$& \textbf{0.138}      & 0.149       & 0.165  & 0.067    \\
        \bottomrule
        \end{tabular}
        \label{s}
    \end{minipage}
\end{table}

\begin{figure}[t]
	\centering
    \includegraphics[width=\linewidth]{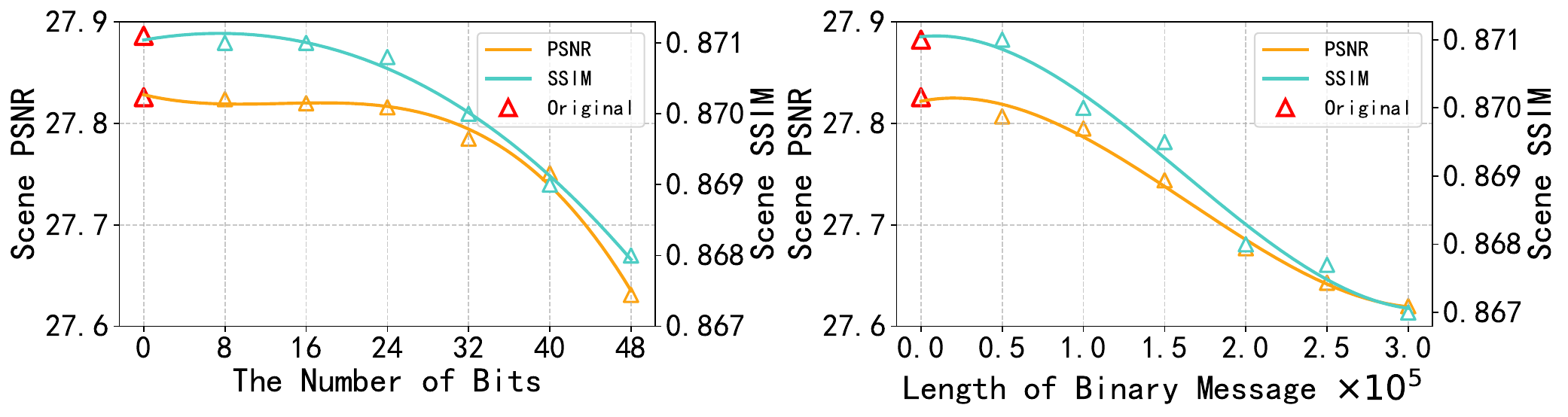} 
	\caption{The interference of watermark capacity on the original scene.}
	\label{fig:tradeoff}
\end{figure}

\paragraph{Capacity vs. Distortion Trade-off.} Figure~\ref{fig:tradeoff} conducted two experiments to demonstrate how increasing watermark payload affects rendering quality:
The left figure is for the number of bits (fixed at 15,000 patches), while the right is for the patch length (48 bits/patch). 
Both used bit-only watermarks. Results show that longer length and more patches lead to greater degradation in rendering quality but the interference remains limited.

\paragraph{Robustness Under Varying Perturbation Strengths.} Table~\ref{fig:crrrrr} presents the model's robustness against watermark attacks of varying intensities. To evaluate this, we subjected the model to cropout attacks with different intensity levels, which shows less than 3\% accuracy drop even with 25\% cropout rate. 
\begin{table}[!ht]
\centering
\scriptsize 
\setlength{\tabcolsep}{3pt} 
\renewcommand{\arraystretch}{1}
\caption{Analysis of robustness against varying Cropout intensities. Despite increasing distortion, the watermark extraction accuracy remains stable.}
\begin{tabular}{@{}lccccc@{}} 
\toprule
Cr & Acc & PSNR(img) & PSNR(obj) & SSIM & LPIPS$\downarrow$ \\
\midrule
5\%   & 100 & 22.611 & 23.009 & 0.835 & 0.139 \\
10\%  & 100 & 22.539 & 23.005 & 0.835 & 0.139 \\
15\%  & 100 & 22.340 & 22.912 & 0.831 & 0.141 \\
20\%  & 98.3 & 22.098 & 22.470 & 0.828 & 0.147 \\
25\%  & 97.7 & 21.563 & 22.133 & 0.824 & 0.155 \\
\bottomrule
\end{tabular}
\label{fig:crrrrr}
\end{table}

\section{Conclusion}
In this paper, we introduce X-SG$^2$S, the first multimodal watermarking framework designed explicitly for 3D Gaussian Splatting. X-SG$^2$S embeds 1-, 2-, or 3-dimensional payloads into a radiance field without altering the original 3DGS parameters or requiring per-scene fine-tuning, thus eliminating time and storage overhead. By pairing a novel modality-specific patchifing method with a lightweight neural policy network, our method adaptively selects the most suitable locations in 3DGS scenes for embedding and reliably recovers the payload after severe 3D corruptions (noise, rotation, translation, cropout). Extensive experiments on diverse  scenes demonstrate a higher payload capacity than existing methods while maintaining visual fidelity indistinguishable from the pristine rendering. We believe X-SG$^2$S will facilitate downstream copyright protection and authentication tasks for rapidly-growing 3DGS assets.

\section*{Acknowledgement}
This research was supported by the National Natural Science Foundation of China (62306117), the Guangdong S\&T Programme (2025B0101120007), the Guangdong Basic and Applied Basic Research Foundation (2026A1515011478), and GJYC program of Guangzhou (2024D03J0005).

%
%
\bibliographystyle{splncs04}
\bibliography{main}

@String(CVPR  = {IEEE Conf. Comput. Vis. Pattern Recog.})

@String(ICCV  = {Int. Conf. Comput. Vis.})

@String(ECCV  = {Eur. Conf. Comput. Vis.})

@String(AAAI  = {AAAI})

@String(TOG   = {ACM Trans. Graph.})

@String(ACMMM = {ACM Int. Conf. Multimedia})

@String(CVPR  = {CVPR})

@String(ICCV  = {ICCV})

@String(ECCV  = {ECCV})

@String(TOG   = {ACM TOG})

@String(ACMMM = {ACM MM})

@inproceedings{chen2025mvsplat,
  title={Mvsplat: Efficient 3d gaussian splatting from sparse multi-view images},
  author={Chen, Yuedong and Xu, Haofei and Zheng, Chuanxia and Zhuang, Bohan and Pollefeys, Marc and Geiger, Andreas and Cham, Tat-Jen and Cai, Jianfei},
  booktitle={European Conference on Computer Vision},
  pages={370--386},
  year={2025},
  organization={Springer}
}

@article{zhang2024gs,
  title={GS-Hider: Hiding Messages into 3D Gaussian Splatting},
  author={Zhang, Xuanyu and Meng, Jiarui and Li, Runyi and Xu, Zhipei and Zhang, Yongbing and Zhang, Jian},
  journal={arXiv preprint arXiv:2405.15118},
  year={2024}
}

@inproceedings{ma2024safe,
  title={Safe-sd: Safe and traceable stable diffusion with text prompt trigger for invisible generative watermarking},
  author={Ma, Zhiyuan and Jia, Guoli and Qi, Biqing and Zhou, Bowen},
  booktitle={Proceedings of the 32nd ACM International Conference on Multimedia},
  pages={7113--7122},
  year={2024}
}

@inproceedings{zhu2024rethinking,
  title={Rethinking Mesh Watermark: Towards Highly Robust and Adaptable Deep 3D Mesh Watermarking},
  author={Zhu, Xingyu and Ye, Guanhui and Luo, Xiapu and Wei, Xuetao},
  booktitle={Proceedings of the AAAI Conference on Artificial Intelligence},
  volume={38},
  number={7},
  pages={7784--7792},
  year={2024}
}

@article{shen2024gamba,
  title={Gamba: Marry gaussian splatting with mamba for single view 3d reconstruction},
  author={Shen, Qiuhong and Wu, Zike and Yi, Xuanyu and Zhou, Pan and Zhang, Hanwang and Yan, Shuicheng and Wang, Xinchao},
  journal={arXiv preprint arXiv:2403.18795},
  year={2024}
}

@inproceedings{shen2021efficient,
  title={Efficient attention: Attention with linear complexities},
  author={Shen, Zhuoran and Zhang, Mingyuan and Zhao, Haiyu and Yi, Shuai and Li, Hongsheng},
  booktitle={Proceedings of the IEEE/CVF winter conference on applications of computer vision},
  pages={3531--3539},
  year={2021}
}

@inproceedings{lee2019set,
  title={Set transformer: A framework for attention-based permutation-invariant neural networks},
  author={Lee, Juho and Lee, Yoonho and Kim, Jungtaek and Kosiorek, Adam and Choi, Seungjin and Teh, Yee Whye},
  booktitle={International conference on machine learning},
  pages={3744--3753},
  year={2019},
  organization={PMLR}
}

@inproceedings{qi2017pointnet,
  title={Pointnet: Deep learning on point sets for 3d classification and segmentation},
  author={Qi, Charles R and Su, Hao and Mo, Kaichun and Guibas, Leonidas J},
  booktitle={Proceedings of the IEEE conference on computer vision and pattern recognition},
  pages={652--660},
  year={2017}
}

@article{kerbl20233d,
  title={3D Gaussian splatting for real-time radiance field rendering.},
  author={Kerbl, Bernhard and Kopanas, Georgios and Leimk{\"u}hler, Thomas and Drettakis, George},
  journal={ACM Trans. Graph.},
  volume={42},
  number={4},
  pages={139--1},
  year={2023}
}

@inproceedings{szymanowicz2024splatter,
  title={Splatter image: Ultra-fast single-view 3d reconstruction},
  author={Szymanowicz, Stanislaw and Rupprecht, Chrisitian and Vedaldi, Andrea},
  booktitle={Proceedings of the IEEE/CVF Conference on Computer Vision and Pattern Recognition},
  pages={10208--10217},
  year={2024}
}

@article{boss2024sf3d,
  title={Sf3d: Stable fast 3d mesh reconstruction with uv-unwrapping and illumination disentanglement},
  author={Boss, Mark and Huang, Zixuan and Vasishta, Aaryaman and Jampani, Varun},
  journal={arXiv preprint arXiv:2408.00653},
  year={2024}
}

@inproceedings{charatan2024pixelsplat,
  title={pixelsplat: 3d gaussian splats from image pairs for scalable generalizable 3d reconstruction},
  author={Charatan, David and Li, Sizhe Lester and Tagliasacchi, Andrea and Sitzmann, Vincent},
  booktitle={Proceedings of the IEEE/CVF Conference on Computer Vision and Pattern Recognition},
  pages={19457--19467},
  year={2024}
}

@inproceedings{ohbuchi2002frequency,
  title={A frequency-domain approach to watermarking 3D shapes},
  author={Ohbuchi, 1 Ryutarou and Mukaiyama, 1 Akio and Takahashi, 2 Shigeo},
  booktitle={Computer graphics forum},
  volume={21},
  number={3},
  pages={373--382},
  year={2002},
  organization={Wiley Online Library}
}

@inproceedings{li2023steganerf,
  title={Steganerf: Embedding invisible information within neural radiance fields},
  author={Li, Chenxin and Feng, Brandon Y and Fan, Zhiwen and Pan, Panwang and Wang, Zhangyang},
  booktitle={Proceedings of the IEEE/CVF International Conference on Computer Vision},
  pages={441--453},
  year={2023}
}

@inproceedings{luo2023copyrnerf,
  title={Copyrnerf: Protecting the copyright of neural radiance fields},
  author={Luo, Ziyuan and Guo, Qing and Cheung, Ka Chun and See, Simon and Wan, Renjie},
  booktitle={Proceedings of the IEEE/CVF International Conference on Computer Vision},
  pages={22401--22411},
  year={2023}
}

@article{li2024gaussianstego,
  title={Gaussianstego: A generalizable stenography pipeline for generative 3d gaussians splatting},
  author={Li, Chenxin and Liu, Hengyu and Fan, Zhiwen and Li, Wuyang and Liu, Yifan and Pan, Panwang and Yuan, Yixuan},
  journal={arXiv preprint arXiv:2407.01301},
  year={2024}
}

@article{ferreira2020robust,
  title={A robust 3D point cloud watermarking method based on the graph Fourier transform},
  author={Ferreira, Felipe ABS and Lima, Juliano B},
  journal={Multimedia Tools and Applications},
  volume={79},
  number={3},
  pages={1921--1950},
  year={2020},
  publisher={Springer}
}

@inproceedings{wang2020logo,
  title={Logo-2k+: A large-scale logo dataset for scalable logo classification},
  author={Wang, Jing and Min, Weiqing and Hou, Sujuan and Ma, Shengnan and Zheng, Yuanjie and Wang, Haishuai and Jiang, Shuqiang},
  booktitle={Proceedings of the AAAI Conference on Artificial Intelligence},
  volume={34},
  number={04},
  pages={6194--6201},
  year={2020}
}

@inproceedings{deitke2023objaverse,
  title={Objaverse: A universe of annotated 3d objects},
  author={Deitke, Matt and Schwenk, Dustin and Salvador, Jordi and Weihs, Luca and Michel, Oscar and VanderBilt, Eli and Schmidt, Ludwig and Ehsani, Kiana and Kembhavi, Aniruddha and Farhadi, Ali},
  booktitle={Proceedings of the IEEE/CVF Conference on Computer Vision and Pattern Recognition},
  pages={13142--13153},
  year={2023}
}

@article{chen2024deep,
  title={Deep compression autoencoder for efficient high-resolution diffusion models},
  author={Chen, Junyu and Cai, Han and Chen, Junsong and Xie, Enze and Yang, Shang and Tang, Haotian and Li, Muyang and Lu, Yao and Han, Song},
  journal={arXiv preprint arXiv:2410.10733},
  year={2024}
}

@article{guo2024splats,
  title={Splats in Splats: Embedding Invisible 3D Watermark within Gaussian Splatting},
  author={Guo, Yijia and Huang, Wenkai and Li, Yang and Li, Gaolei and Zhang, Hang and Hu, Liwen and Li, Jianhua and Huang, Tiejun and Ma, Lei},
  journal={arXiv preprint arXiv:2412.03121},
  year={2024}
}

@article{huang2024gaussianmarker,
  title={Gaussianmarker: Uncertainty-aware copyright protection of 3d gaussian splatting},
  author={Huang, Xiufeng and Li, Ruiqi and Cheung, Yiu-ming and Cheung, Ka Chun and See, Simon and Wan, Renjie},
  journal={Advances in Neural Information Processing Systems},
  volume={37},
  pages={33037--33060},
  year={2024}
}

@inproceedings{jang20253d,
  title={3d-gsw: 3d gaussian splatting for robust watermarking},
  author={Jang, Youngdong and Park, Hyunje and Yang, Feng and Ko, Heeju and Choo, Euijin and Kim, Sangpil},
  booktitle={Proceedings of the Computer Vision and Pattern Recognition Conference},
  pages={5938--5948},
  year={2025}
}

@inproceedings{chen2025guardsplat,
  title={GuardSplat: Efficient and Robust Watermarking for 3D Gaussian Splatting},
  author={Chen, Zixuan and Wang, Guangcong and Zhu, Jiahao and Lai, Jianhuang and Xie, Xiaohua},
  booktitle={Proceedings of the Computer Vision and Pattern Recognition Conference},
  pages={16325--16335},
  year={2025}
}

@inproceedings{zhu2018hidden,
  title={Hidden: Hiding data with deep networks},
  author={Zhu, Jiren and Kaplan, Russell and Johnson, Justin and Fei-Fei, Li},
  booktitle={Proceedings of the European conference on computer vision (ECCV)},
  pages={657--672},
  year={2018}
}

@inproceedings{jing2021hinet,
  title={Hinet: Deep image hiding by invertible network},
  author={Jing, Junpeng and Deng, Xin and Xu, Mai and Wang, Jianyi and Guan, Zhenyu},
  booktitle={Proceedings of the IEEE/CVF international conference on computer vision},
  pages={4733--4742},
  year={2021}
}

@article{fernandez2024video,
  title={Video seal: Open and efficient video watermarking},
  author={Fernandez, Pierre and Elsahar, Hady and Yalniz, I Zeki and Mourachko, Alexandre},
  journal={arXiv preprint arXiv:2412.09492},
  year={2024}
}

@article{hu2025videomark,
  title={Videomark: A distortion-free robust watermarking framework for video diffusion models},
  author={Hu, Xuming and Li, Hanqian and Li, Jungang and Huang, Yu and Liu, Aiwei},
  journal={arXiv preprint arXiv:2504.16359},
  year={2025}
}

@inproceedings{zhang2018unreasonable,
  title={The unreasonable effectiveness of deep features as a perceptual metric},
  author={Zhang, Richard and Isola, Phillip and Efros, Alexei A and Shechtman, Eli and Wang, Oliver},
  booktitle={Proceedings of the IEEE conference on computer vision and pattern recognition},
  pages={586--595},
  year={2018}
}

@article{wang2004image,
  title={Image quality assessment: from error visibility to structural similarity},
  author={Wang, Zhou and Bovik, Alan C and Sheikh, Hamid R and Simoncelli, Eero P},
  journal={IEEE transactions on image processing},
  volume={13},
  number={4},
  pages={600--612},
  year={2004},
  publisher={IEEE}
}

@article{mildenhall2021nerf,
  title={Nerf: Representing scenes as neural radiance fields for view synthesis},
  author={Mildenhall, Ben and Srinivasan, Pratul P and Tancik, Matthew and Barron, Jonathan T and Ramamoorthi, Ravi and Ng, Ren},
  journal={Communications of the ACM},
  volume={65},
  number={1},
  pages={99--106},
  year={2021},
  publisher={ACM New York, NY, USA}
}

@article{mildenhall2019local,
  title={Local light field fusion: Practical view synthesis with prescriptive sampling guidelines},
  author={Mildenhall, Ben and Srinivasan, Pratul P and Ortiz-Cayon, Rodrigo and Kalantari, Nima Khademi and Ramamoorthi, Ravi and Ng, Ren and Kar, Abhishek},
  journal={ACM Transactions on Graphics (ToG)},
  volume={38},
  number={4},
  pages={1--14},
  year={2019},
  publisher={ACM New York, NY, USA}
}

@article{Liu2020InfiniteNP,
  title={Infinite Nature: Perpetual View Generation of Natural Scenes from a Single Image},
  author={Andrew Liu and Richard Tucker and V. Jampani and Ameesh Makadia and Noah Snavely and Angjoo Kanazawa},
  journal={2021 IEEE/CVF International Conference on Computer Vision (ICCV)},
  year={2020},
  pages={14438-14447},
  url={https://api.semanticscholar.org/CorpusID:229297871}
}

@article{Zhou2024DiffGSFG,
  title={DiffGS: Functional Gaussian Splatting Diffusion},
  author={Junsheng Zhou and Weiqi Zhang and Yu-Shen Liu},
  journal={ArXiv},
  year={2024},
  volume={abs/2410.19657},
  url={https://api.semanticscholar.org/CorpusID:273638402}
}

@inproceedings{barron2022mip,
  title={Mip-nerf 360: Unbounded anti-aliased neural radiance fields},
  author={Barron, Jonathan T and Mildenhall, Ben and Verbin, Dor and Srinivasan, Pratul P and Hedman, Peter},
  booktitle={Proceedings of the IEEE/CVF conference on computer vision and pattern recognition},
  pages={5470--5479},
  year={2022}
}

@misc{li2026vlaattcadaptivetesttimecompute,
      title={VLA-ATTC: Adaptive Test-Time Compute for VLA Models with Relative Action Critic Model}, 
      author={Wenhao Li and Xiu Su and Dan Niu and Yichao Cao and Hongyan Xu and Zhe Qu and Lei Fan and Shan You and Chang Xu},
      year={2026},
      eprint={2605.01194},
      archivePrefix={arXiv},
      primaryClass={cs.RO},
      url={https://arxiv.org/abs/2605.01194}, 
}

@misc{li2026sentinelvlametacognitivevlamodel,
      title={Sentinel-VLA: A Metacognitive VLA Model with Active Status Monitoring for Dynamic Reasoning and Error Recovery}, 
      author={Wenhao Li and Xiu Su and Yichao Cao and Hongyan Xu and Xiaobo Xia and Shan You and Yi Chen and Chang Xu},
      year={2026},
      eprint={2605.01191},
      archivePrefix={arXiv},
      primaryClass={cs.RO},
      url={https://arxiv.org/abs/2605.01191}, 
}

@article{fang2025your,
  title={Your data is not perfect: Towards cross-domain out-of-distribution detection in class-imbalanced data},
  author={Fang, Xiang and Easwaran, Arvind and Genest, Blaise and Suganthan, Ponnuthurai Nagaratnam},
  journal={Expert Systems with Applications},
  year={2025}
}

@inproceedings{cvpr/gait3d_v1,
  author    = {Jinkai Zheng and
               Xinchen Liu and
               Wu Liu and
               Lingxiao He and
               Chenggang Yan and
               Tao Mei},
  title     = {Gait Recognition in the Wild with Dense 3D Representations and {A}
               Benchmark},
  booktitle = CVPR,
  pages     = {20228--20237},
  year      = {2022}
}

@inproceedings{parsinggait_gps,
  author       = {Jinkai Zheng and
                  Xinchen Liu and
                  Shuai Wang and
                  Lihao Wang and
                  Chenggang Yan and
                  Wu Liu},
  title        = {Parsing is All You Need for Accurate Gait Recognition in the Wild},
  booktitle    = ACMMM,
  pages        = {116--124},
  year         = {2023}
}

@article{He2023UnifiedSGGHOI,
  author  = {Tao He and Lianli Gao and Jingkuan Song and Yuan-Fang Li},
  title   = {Toward a Unified Transformer-based Framework for Scene Graph Generation and Human-Object Interaction Detection},
  journal = {IEEE Transactions on Image Processing},
  volume  = {32},
  pages   = {6274--6288},
  year    = {2023}
}

@inproceedings{He2022OpenVocabularySGG,
  author    = {Tao He and Lianli Gao and Jingkuan Song and Yuan-Fang Li},
  title     = {Towards Open-Vocabulary Scene Graph Generation with Prompt-Based Finetuning},
  booktitle = {European Conference on Computer Vision (ECCV)},
  pages     = {56--73},
  year      = {2022}
}
\end{document}